\documentclass{edp-jp4}
\usepackage{graphicx}

\begin{document}

\title{Social inertia and diversity in collaboration networks} 
\author{Jos\'e J. Ramasco}\address{Physics Department, Emory University,
Atlanta, Georgia 30322, USA}

\maketitle

\begin{abstract} 

Random graphs are useful tools to study social interactions. In particular, 
the use of weighted random graphs allows to handle a high level of information
concerning which agents interact and in which degree the interactions
take place. Taking advantage of this representation, we recently defined a
magnitude, the Social Inertia, that measures the eagerness of agents to keep
ties with previous partners. To study this magnitude, we used collaboration
networks that are specially appropriate to obtain valid 
statistical results 
due to the large size of publically available databases. In this work, I 
study the Social Inertia in two of these 
empirical
networks, IMDB movie database and condmat. More specifically, I 
focus on how the Inertia relates 
to other properties of the graphs, and show that the Inertia provides 
information on how the
weight of neighboring edges correlates. A social interpretation of this 
effect is also offered.

\end{abstract}

\section{Introduction}

The theory of complex networks has recently produced a great deal of
interest in a very multidisciplinary community (for recent reviews on the
field see \cite{barabasi02,sergei03,romu04,newman03}). It has been applied
with success to a number of fields spanning from the Internet and the 
World-Wide Web \cite{barabasi99,albert99,romu01} to protein interactions in
cells \cite{jeong00,schwikowski00,wuchty03}. The study of human society 
is another topic where networks can play an important role. In this 
particular case, the
vertices represent individuals and the edges social interactions
such as professional, friendship, or family relationships. These interactions
can appear on different levels of intensity. How strength our friendship with
other person is cannot be seen as a white-and-black concept but as a full scale
of colors. This means that the best 
networks to
characterize social interactions are weighted graphs. Weighted graphs
include a magnitude associated to the edges, a so-called {\it weight},
that accounts for the quality of each connection \cite{yook01}.  Here I will 
apply the mathematical tools designed for weighted graphs to collaborations 
networks.

So far the major problem for the study of social 
networks has been the absence of large enough databases from which
reliable statistical conclusions could be extracted. However, on the edge of the
digital era, this 
restriction does no longer exist for a particular kind of social networks, the
so called {\it collaboration networks}. This type of networks are obtained 
from public databases containing artistic or scientific productions 
such as books, movies or papers, together with the names of the people 
authoring those works. The network is then formed by connecting together 
pairs of persons who have co-authored a common work 
\cite{barabasi99,newman01b}. This graph is 
undirected, the relations are reciprocal, and it may be weighted. The weights
can be used to represent how many times a certain partnership has
taken place, maintaining thus a high degree of information in a single
graph \cite{newman01b,barrat04}. Recently, we exploited the information
contained in the weights of the links to define a new quantity, the {\it Social
Inertia}, which measures the eagerness of the authors to keep working over and
over with the same team \cite{ramasco06a}. My intention in this paper is to 
study in detail the
foundations of this magnitude and to show why it gives new
information different from previous metrics. In order to illustrate these 
points, I will use networks obtained from the IMDB movie database
\cite{barabasi99,note1} and from
condmat \cite{newman01b}. The IMDB database comprehends $383640$ actors and
$127823$ movies, while the data from condmat contains $16721$ authors and
$22002$ papers.

\section{Social inertia}

Let us start by considering a network where the nodes are authors or actors, the
edges represent partnerships and the weight of the edges, $w_{ij}$ for a link
between $i$ and $j$, the number
of times a co-authorship between authors $i$ and $j$ has been repeated. The
degree of a node $i$, the number of connections $k_i$, corresponds to the number
of different coauthors a particular actor has had. Another important magnitude
is the strength $s_i$, which is the sum over all the weights of 
the links of node $i$. In our
case, $s_i$ is the total number of partnerships $i$ had. The social inertia 
for $i$ is then defined as
\begin{equation}
\mathcal{I}_i = s_i/k_i ,
\end{equation}
and accounts for how many of the partnerships have taken place with different
partners. $\mathcal{I}_i$ measures the level of conservatism of $i$, how open he
or she is to collaborate with different people. Its limits are $\mathcal{I}_i
\to 1$ if the actor has never repeated collaborators, and $\mathcal{I}_i \to
q_i$ if all her works were carried out with the same team, where $q_i$ stands
for the total experience of $i$ (number of works she has authored). 

\section{Relation between the Inertia and other properties of networks}

\begin{figure}
\includegraphics[width=7cm]{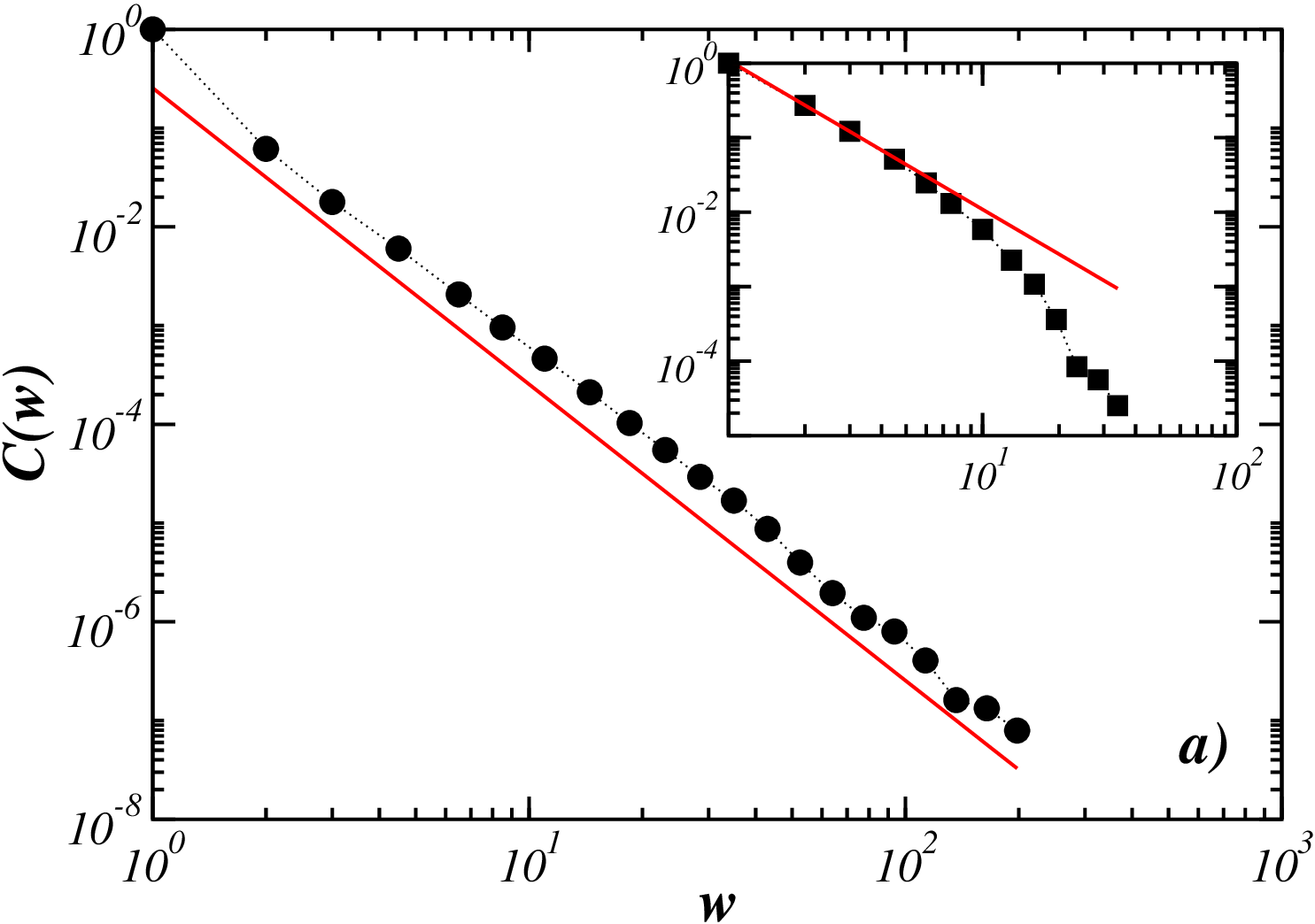}
\qquad
\includegraphics[width=7cm]{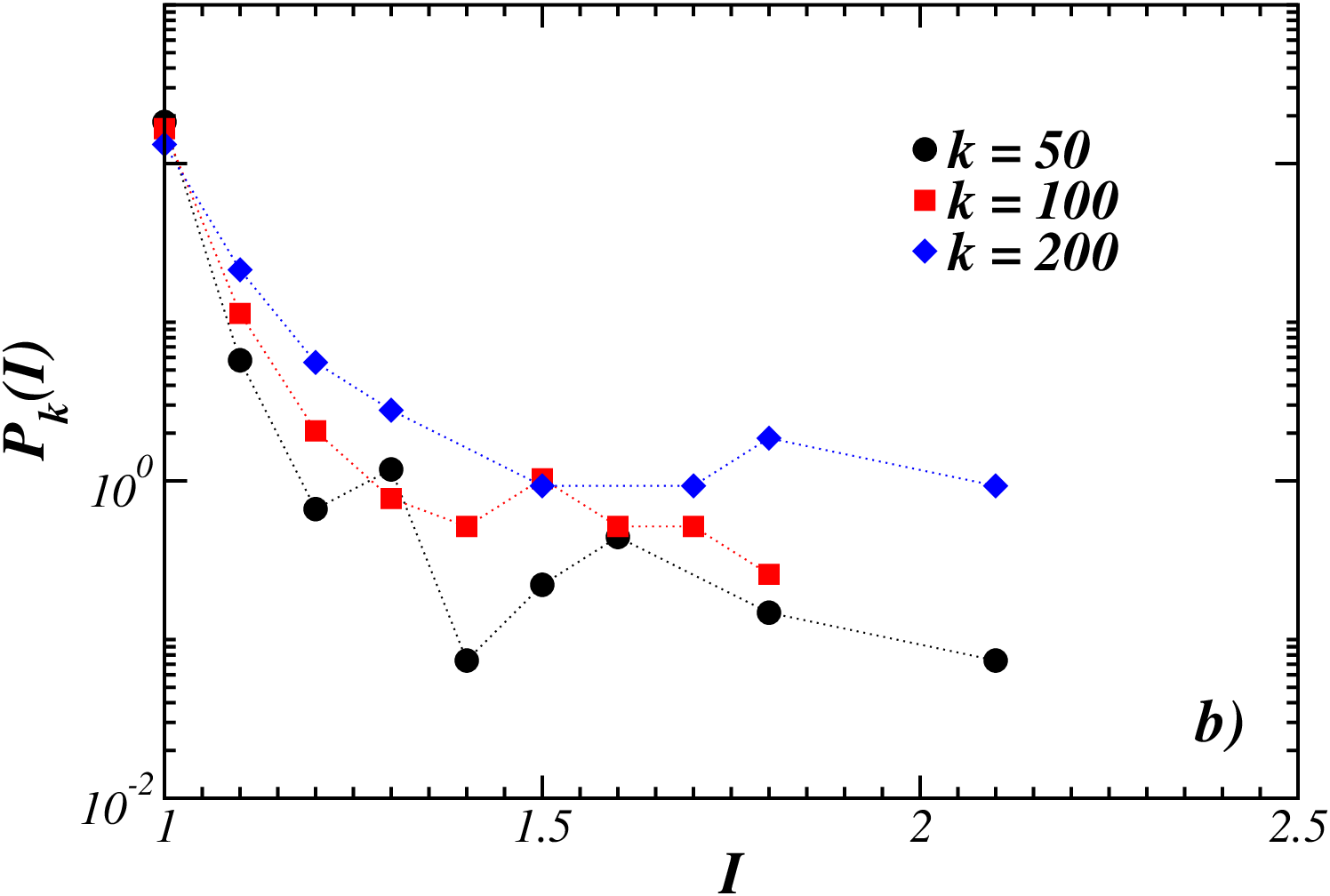}
\caption{a) shows the cumulative distribution of weights, $C(w)$, for the
movie network (main plot) and condmat (inset). The red line in the main plot
corresponds to a power law with exponent $-3$ and the one in the inset to an
exponent $-2$ (the limit to have finite second moment). The b) plot contains the
distribution of values of Inertia for nodes with a given value of $k$ ($50, 100,
200$) in the actor network.}
\end{figure}

The Inertia is the average weight of the links of a node, 
$\mathcal{I}_i = s_i/k_i = (1/k_i) \sum_{j \in \nu(j)} w_{ij}$ where $\nu(i)$
represents the set of $k_i$ neighbors of $i$. If we
consider a network where all the weights are alike, the Inertia is a constant.
Unweighted graphs are a particular case of this situation with $\mathcal{I}_i =
1$ for all $i$. If there exist a weight distribution in the graph 
$P_w(w)$, then the values allowed to $\mathcal{I}$
depend on how wide such distribution is. For distributions with a finite
second moment and for nodes with high degrees $k$, the Central Limit Theorem 
implies 
that their strengths must show a Gaussian distribution around a central value
$\langle s \rangle (k)$ and that the deviation of this distribution should 
grows as $\sigma_s(k) \sim k^{1/2}$ with the degree. This leads in turn to a 
Gaussian distribution of the
fluctuations of the
Inertia of nodes with the same degree $k$, with the standard
deviation decreasing with the degree as $\sigma_\mathcal{I} (k) \sim
k^{-1/2}$. In other words: the Inertia should be better and better determined,
the larger the degree of a node becomes. If the degree of a node is known, 
there remains almost no uncertainty in its value of the Inertia (specially if its
degree is high). This argument seems to establish that the Inertia is a
magnitude dependent of others as the degree, but is it really like
that in real-world networks?

In order to give an answer to this question, I have plotted in Figure 1a 
the cumulative weight 
distribution 
($C(w) = \int_w^\infty dw' P(w')$) for both empirical networks (actors 
and condamt). For the two examples, the weight distribution is
wide but decays faster than $C(w) \sim w^{-2}$, which implies that these
distributions 
have finite second moments. However, as can be observed in Fig. 1b, the
distribution of values of inertia for nodes with a given value of the degree, 
$P_k(\mathcal{I})$, does not tend to a Gaussian form for high values of $k$.
Otherwise, the curves in Fig. 1b should tend to a parabola when $k$ increases.
This fact is in contradiction with the argument above. Another point of 
conflict is its final
prediction for the Inertia: the deviation of the values of $\mathcal{I}$ 
for nodes with a certain degree $k$,
$\sigma_\mathcal{I}(k)$, does not decay as $k^{-1/2}$ for any of the networks
studied. Instead, it grows for the actor network, see Figs. 2a, and remains
almost constant for the condamt (Fig. 2b). This leads
to a kind of indetermination rule: for the actors, the higher the degree of a 
node is, the less we know a priori about its possible value of the Inertia. And
for the condmat, knowing the degree does not tell us anything about the Inertia.
The values of the average 
Inertia
as a function of the degree is also represented in the same Figures and, 
in contrast to what happens in transport
networks \cite{barrat04}, it does not change significantly.

\begin{figure}
\includegraphics[width=7cm]{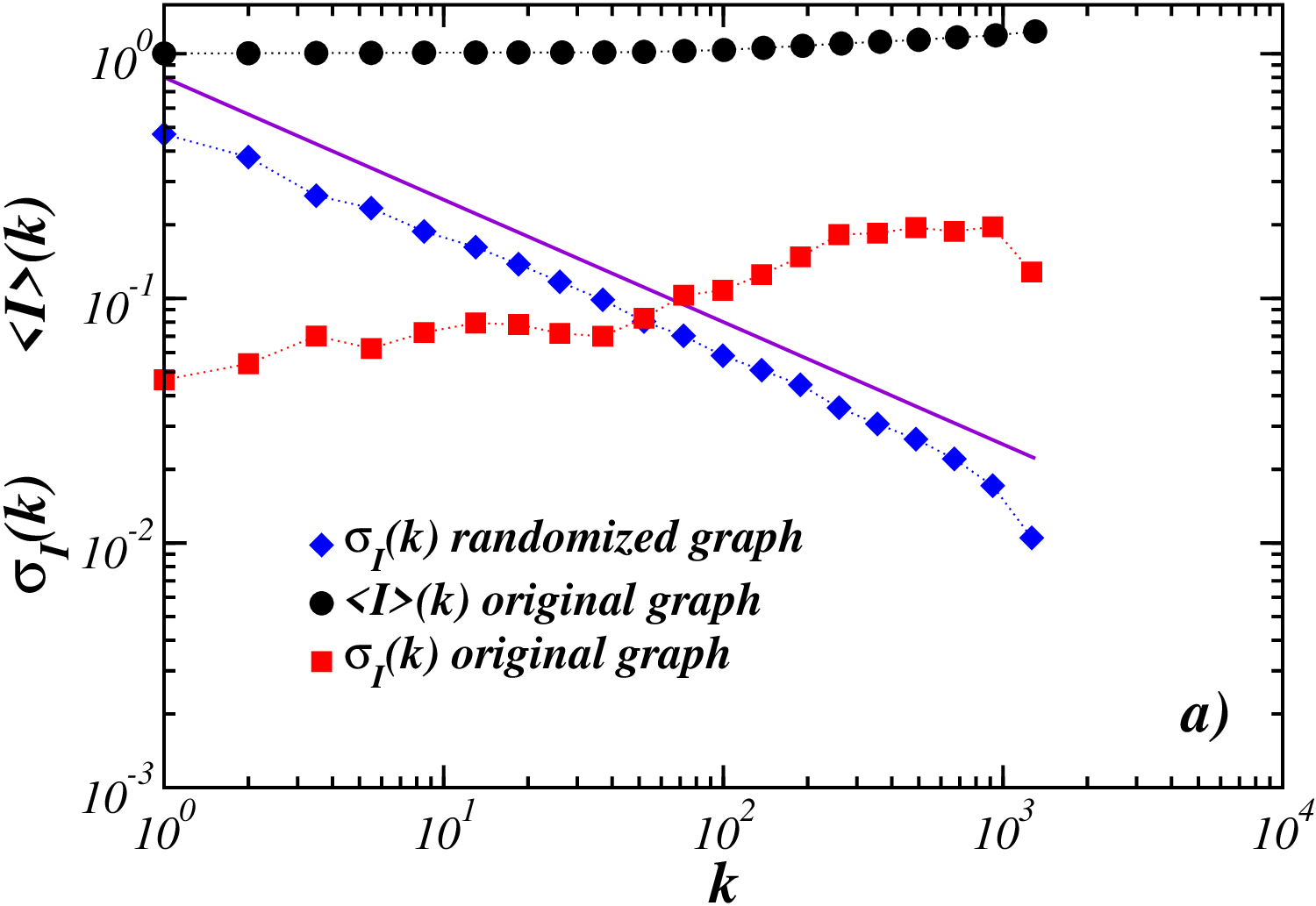}
\qquad
\includegraphics[width=7cm]{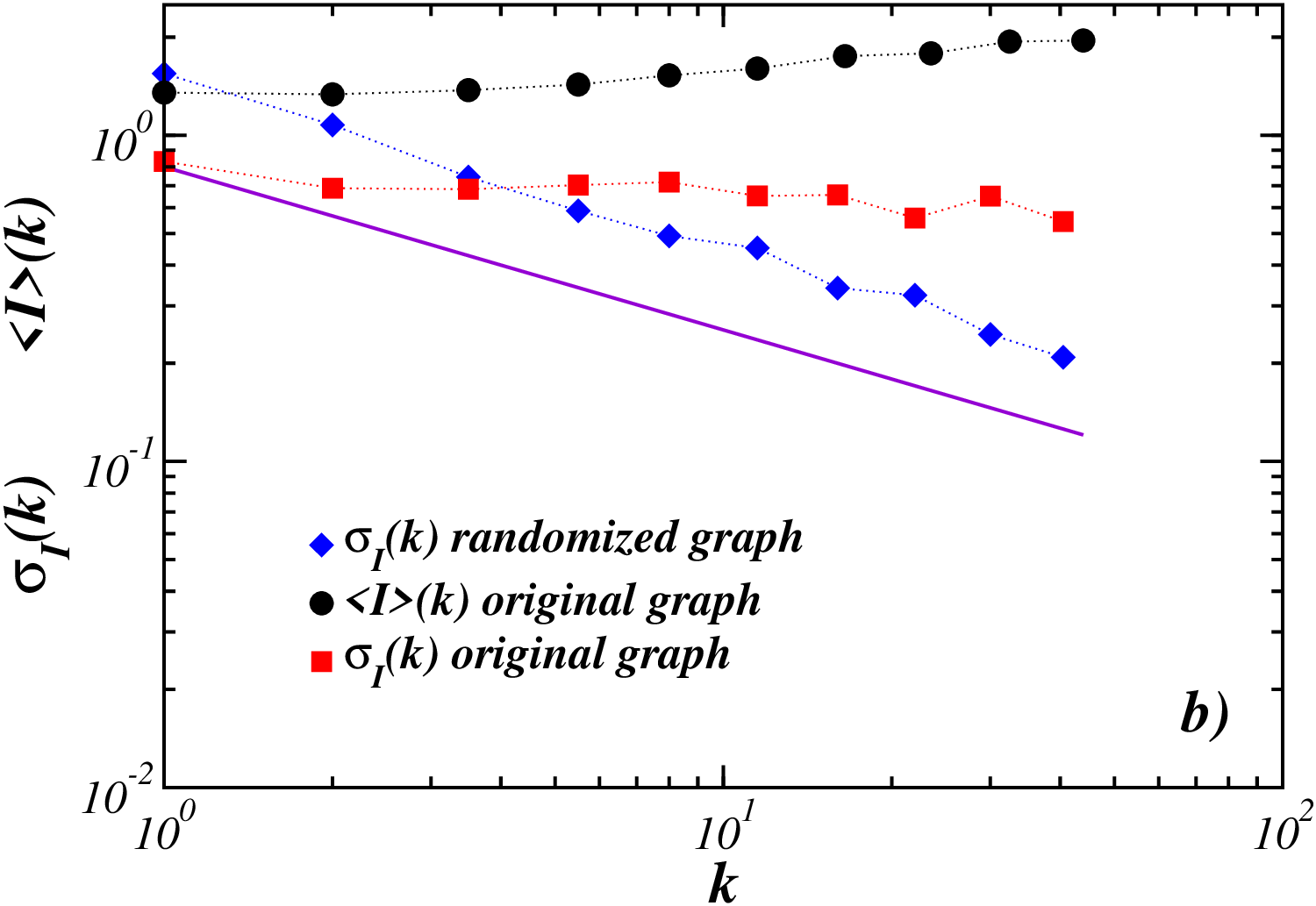}
\caption{Average Inertia and standard deviation as functions of the degree. The
data in a) are for the actor network and those in b) for the condmat.
The blue diamonds correspond to the randomized networks obtained switching
the values of the weights of the links of the original networks (see explanation
in the text). The two straight lines represent the predicted $k^{-1/2}$ decay
for uncorrelated networks.}
\end{figure}

One may wonder then what these networks have in particular to show this 
behavior. The answer is profusely discussed in Ref. \cite{ramasco06b} and is
related to the fact that the weights of the edges are not
randomly distributed. The edges of a node tend in general to
be uniform, more than in a purely random distribution. These correlations 
imply that nodes with 
the same degree can have very different values of the strength and 
consequently vary in the Inertia. Weak links are concentrated in 
certain areas of the network and the same happens 
with the strong links. To illustrate this mechanism, I have disordered the 
weights
of the links: maintaining the same topological structure of the network, the 
weight of each link is interchanged with that of another randomly 
chosen edge. The effect on $\sigma_\mathcal{I}(k)$ can be seen in Figs. 2a and
2b. For the randomized networks, the deviation decays as $k^{-1/2}$ following
the prediction done by the argument discussed above for uncorrelated graphs. 

From the social perspective, this effect means that
the authors or actors display a tendency towards keeping their partnerships in
relative similar levels. Some people are quite faithful and go on repeating 
with the same
collaborators, others change of collaborators with high frequency and do not
maintain a partnership for very long. These
are the two extremes but of course there is a full scale of
behaviors for the agents in the middle. However, extreme conducts are here more likely than 
in a completely random situation.
Consequently, even if 
the number of 
different partners is the same for two actors, it is not
easy to predict to which category they 
belong. The difficulty of doing so may even increase with an increasing number of
partners.

\section{Conclusions}

In summary, I have studied here how the Inertia, the average weight, of the 
nodes relates to other magnitudes in social networks. A very simple theoretical 
arguments suggests that knowing a certain magnitude as the degree, one has 
the Inertia of a node specified. I have checked the validity of 
this argument in two real-world social networks: the IMDB movie database and 
the condmat. Both of
these cases show that the theoretical prediction fails. The reason for the 
failure is the 
existence of weight-weight correlations in real networks. This fact implies
that the distribution of the Inertia contains important information on the
behavior of the agents. From a social point 
of view, the existence of these 
correlation indicate the presence of two different type of behaviors. Some
agents are faithful to their partners and maintain in average a high level of
collaboration with them, while others have a tendency to change quickly their 
collaborators without allowing the partnerships to go too far.


\begin{thebibliography}{99}

\bibitem{barabasi02}
R. Albert and A.-L. Barab{\'a}si, Rev. Mod. Phys. {\bf 74},  47  (2002).

\bibitem{sergei03}
S.N. Dorogovtsev and J.F.F. Mendes, {\em Evolution of networks: From
  Biological Nets to the Internet and WWW} (Oxford University Press,
  Oxford, 2003).

\bibitem{romu04}R. Pastor-Satorras and A. Vespignani, 
{\em Evolution and structure of the
  Internet: A statistical physics approach} (Cambridge University Press,
  Cambridge, 2004).


\bibitem{newman03}
M.E.J. Newman, SIAM Review {\bf 45},  167  (2003).

\bibitem{barabasi99}
A.-L. Barab\'asi and R. Albert, Science {\bf 286}, 509 (1999).

\bibitem{albert99}
R. Albert, H. Jeong, and A.-L. Barab{\'a}si, Nature {\bf 401},  130  (1999).

\bibitem{romu01}
R. Pastor-Satorras, A. V{\'a}zquez, and A. Vespignani,
Phys. Rev. Lett. {\bf 87},  258701  (2001).

\bibitem{jeong00}
H. Jeong, B. Tombor, R. Albert, Z.N. Oltvai, and A.-L. Barab\'asi,
Nature {\bf 407}, 651 (2000).

\bibitem{schwikowski00}
B. Schwikowski, P. Uetz, and S. Fields, Nat. Biotech. {\bf 18}, 1257 (2000).

\bibitem{wuchty03}
S. Wuchty, Z.N. Oltvai, and A.-L. Barab\'asi, Nat. Genet. {\bf 35}, 176 (2003). 

\bibitem{yook01}
S.H. Yook, H. Jeong, A.-L. Barab\'asi, and Y. Tu, Phys. Rev. Lett. {\bf 86}, 5835
(2001).

\bibitem{newman01b}
M.E.J. Newman, Proc. Natl. Acad. Sci. USA {\bf 98}, 404 (2001); Phys. Rev. E
{\bf 64}, 016131 and 016132 (2001).

\bibitem{barrat04}
A. Barrat, M. Barth\'elemy, R. Pastor-Satorras, and A. Vespignani, Proc. Natl.
Acad. Sci. USA {\bf 101}, 3747 (2004).

\bibitem{ramasco06a} 
J.J. Ramasco and S.A. Morris, Phys. Rev. E {\bf 73}, 016122 (2006).

\bibitem{note1} Data available at
\texttt{http://www.nd.edu/$\sim$networks/dat\-ab\-a\-s\-e/index.html}

\bibitem{mariangels06} 
M.A. Serrano, M. Bogu\~n\'a, and R. Pastor-Satorras, cond-mat/0609029 (2006).

\bibitem{ramasco06b} 
J.J. Ramasco and B. Gon\c calves, cond-mat/0609776 (2006).


\end{thebibliography}
\end{document}